\documentclass{astlb}

\usepackage[utf8]{inputenc}
\usepackage[english]{babel}

\usepackage{caption}
\usepackage{color}

\usepackage{amsmath}
\usepackage{amssymb}

\usepackage{graphicx}
\usepackage{multirow}
\usepackage{rotating}
\usepackage{lscape}
\usepackage{caption}
\usepackage{subcaption}
\usepackage{nameref}
\usepackage{fancyref}
\usepackage{mathtools}
\usepackage{gensymb}
\usepackage{array}
\usepackage{tabularx}
\usepackage{adjustbox}
\usepackage[hyperfootnotes=false]{hyperref}

\newcolumntype{C}{>{\small\centering\arraybackslash}X}

\newcommand\ax{\mbox{AX~J1745.6--2901}}
\newcommand\sgr{\mbox{SGR~J1745--2900}}
\newcommand\cxo{\mbox{CXOGC~J174540.0-290005}}
\newcommand\nus{\textit{NuSTAR}}
\newcommand\fpma{FPMA}
\newcommand\fpmb{FPMB}
\newcommand\swift{\textit{Swift}}
\newcommand{\sgra}{Sgr~A$^{\star}$}

\def\arcsec{\hbox{$^{\prime\prime}$}}

\def\*{$^{*}$}
\def\ц│{$^{\mbox{\small a}}$}
\def\ц┌{$^{\mbox{\small b}}$}
\def\ц≈{$^{\mbox{\small c}}$}
\def\ц┤{$^{\mbox{\small d}}$}
\def\ц└{$^{\mbox{\small e}}$}
\def\etal{{et~al.}}
\sloppypar

\begin{document}

\journalinfo{2021}{47}{4}{214}[223]

\title{Phase-Resolved Spectroscopy of the Magnetar \sgr\ Based \\ on Data from the \nus\ Observatory}

\author{\bf E.A.~Kuznetsova\email{eakuznetsova@cosmos.ru}\address{1}, A.A.~Lutovinov\address{1}, A.N.~Semena\address{1}
\addresstext{1}{Space Research Institute, Russian Academy of Sciences, Profsoyuznaya ul. 84/32, Moscow, 117997 Russia}}

% \affil{
% {\it Space Research Institute of the Russian Academy of Sciences, Moscow}$^1$}

\shortauthor{Kuznetsova \etal}

\shorttitle{PHASE-RESOLVED SPECTROSCOPY OF \sgr} 
\submitted{29.12.2020}

% \vspace{2mm}
% \received{29.12.2020}
%\received{\today}
% \revised{02.02.2021}

% \sloppypar
% \vspace{2mm}

\begin{abstract}
\noindent
The magnetar \sgr\ located in the vicinity of the supermassive black hole Sgr~A$^{\star}$ was detected during its X-ray outburst with the \swift/XRT telescope in April 2013. For several months after its detection the source was observed with the \nus\ observatory, which allowed pulsations with a period $\sim3.76$~s to be recorded. Using these observations, we have studied in detail the dependence of the pulse profile and the pulsed fraction on the energy and intensity of the magnetar. The pulsed fraction in the 3--5 and 5--10 keV energy bands is shown to be 40--50\%, slightly increasing with decreasing flux. We have performed phase-resolved spectroscopy for the source in the energy band from 3 to $\sim$40~keV and show that the temperature of the emitting regions remains fairly stable during the pulse, while their apparent size changes significantly with phase.

\end{abstract}

% \noindent
%{\bf PACS codes:\/} ?????

% \vfill
% \noindent\rule{8cm}{1pt}\\
% {$^*$ E-mail $<$a.shtykovsky@iki.rssi.ru$>$}

% \clearpage

%***************************************************************
\section*{Introduction}

The class of magnetars, which are most unpredictable in their behavior, stands out among the great number of neutron stars. Magnetars are isolated neutron stars with strong magnetic fields up to $B\sim10^{14}-10^{15}$~G that are the energy source of these stars. They manifest themselves as X-ray pulsars with periods $P\simeq(0.3-12)$~s and spin-down rates $\dot{P}\simeq(10^{-15}-10^{-11})$~s~s$^{-1}$. Currently, there are 30 known magnetars \footnote{The online magnetar catalog is accessible at \url{http://www.physics.mcgill.ca/~pulsar/magnetar/main.html}, 24 confirmed sources, and 6 candidates \citep{Olausen14}}.

Magnetars are sources of persistent X-ray emission consisting of two components: thermal, which can be represented as blackbody radiation with a temperature $kT\sim0.3-0.5$~keV, and non-thermal, which is described by a power-law with a photon index $\Gamma\sim2-4$ \citep{Kaspi17}. Apart from persistent emission, powerful X-ray outbursts with luminosities reaching $L_{X}\sim10^{47}$~erg~s$^{-1}$ and durations from fractions to hundreds of seconds can be recorded from magnetars \citep{Turolla15}. Apart from bright and short outbursts, a significant increase in persistent flux accompanied by a succeeding slow decrease to the initial level, which can last from months to several years, is also observed from magnetars \citep[for a review, see][]{Turolla15, Kaspi17}. The manifestations of outbursting activity by magnetars can presumably be caused by neutron star crust deformations, the so-called starquakes.

\begin{table*}
\noindent
\begin{center}
\caption{The \nus\ observations used in this paper. The MJD epochs are specified for the beginning of the observations}
\vspace{2mm}
\begin{tabular}{c|c|c|c}
\hline
  ObsID                  & Date          & MJD      & Exposure time\\
  \hline
  30001002006            & Apr. 26, 2013 & 56408.1  & 37,1~ks    \\ 
  80002013002            & Apr. 27, 2013 & 56409.3  & 49,7~ks    \\ 
  80002013004            & May 4, 2013   & 56416.7  & 38,5~ks    \\ 
  80002013006            & May 11, 2013  & 56423.6  & 32,6~ks    \\ 
  80002013012            & June 14, 2013 & 56457.4  & 26,8~ks    \\ 
  80002013014$^{\star}$  & July 7, 2013  & 56480.2  & 8,6~ks     \\
  80002013016$^{\star}$  & July 7, 2013  & 56480.5  & 21,0~ks    \\[1mm]
\hline
\end{tabular}
\label{tab:obs}
\end{center}
\small{$^{\star}$ The observations for which only the \fpma\ data were used, because the \fpmb\ data were contaminated by stray light from an unknown source.}
\end{table*}

The source \sgr\ is one of the representatives of magnetars. An X-ray flare from an unknown source was detected on April 24, 2013, during a regular monitoring of the Galactic center with the Burst Alert Telescope (BAT) onboard the N.~Gehrels \swift\ space observatory \citep{Degenaar13}, from which a short ($\sim32$~ms) X-ray burst was detected a day later \citep{Kennea13a}. This event served as a trigger for a series of \swift/XRT observations from which the source was found to be spatially unresolvable with the supermassive black hole (SMBH) Sagittarius~A$^{\star}$ (hereafter Sgr~A$^{\star}$) located at the center of our Galaxy \citep{Kennea13b}. Later on, based on data from the {\it Chandra} observatory, \cite{Rea13} resolved the sources \sgr\ and Sgr~A$^{\star}$ determining the angular distance between them, 2\farcs4. Observations of \sgr\ with the \nus\ observatory revealed pulsations with a period $P\sim3.76$~s and a spin-down rate $\dot{P}\sim6.5\times10^{-12}$~s~s$^{-1}$ \citep{Mori13,Kaspi14}. Assuming \sgr\ to be a magnetic dipole in a vacuum, \cite{Mori13} estimated the magnetic field $B\sim1.6\times10^{14}$~G, spin-down power $\dot{E}\simeq5\times10^{33}$~erg~s$^{-1}$, and characteristic age $P/2\dot{P}\simeq9\times10^3$~yr. Similar estimates of the timing parameters were obtained using observations with other telescopes in both X-ray and radio bands \citep{Rea13,Shannon13,Coti15,Coti17,Lynch15,Pennucci15}. It was shown that the spectrum of the persistent emission from the magnetar could be represented as a combination of blackbody radiation with a temperature $kT\sim1$~keV and a power-law with a photon index $\Gamma\sim1.5$ \citep[see, for e.g.][]{Mori13}. A long-term monitoring of \sgr\ with the \nus\ and {\it Chandra} observatories revealed a monotonic decrease in the flux from the magnetar and the temperature of the emitting region $kT$ \citep{Kaspi14,Coti15,Coti17,Rea20}.

In this paper we present the results of our timing analysis (the pulse profiles and the pulsed fraction) and phase-resolved spectroscopy for the magnetar \sgr\ based on data from the \nus\ observatory for several months after its X-ray outburst occurred in April 2013.
%***************************************************************
\section*{Observations and data reduction}
\label{sec:obs}

After the April 24, 2013 outburst, a four-month-long program of observations of the magnetar \sgr\ with the \nus\ space observatory \citep{Harrison13} was carried out from April 26 to August 13, 2013. The NuSTAR observatory consists of two telescope modules, \fpma\ and \fpmb, with the operating energy band 3--79~keV.

\begin{figure*}
\centering
\includegraphics[width=0.45\textwidth]{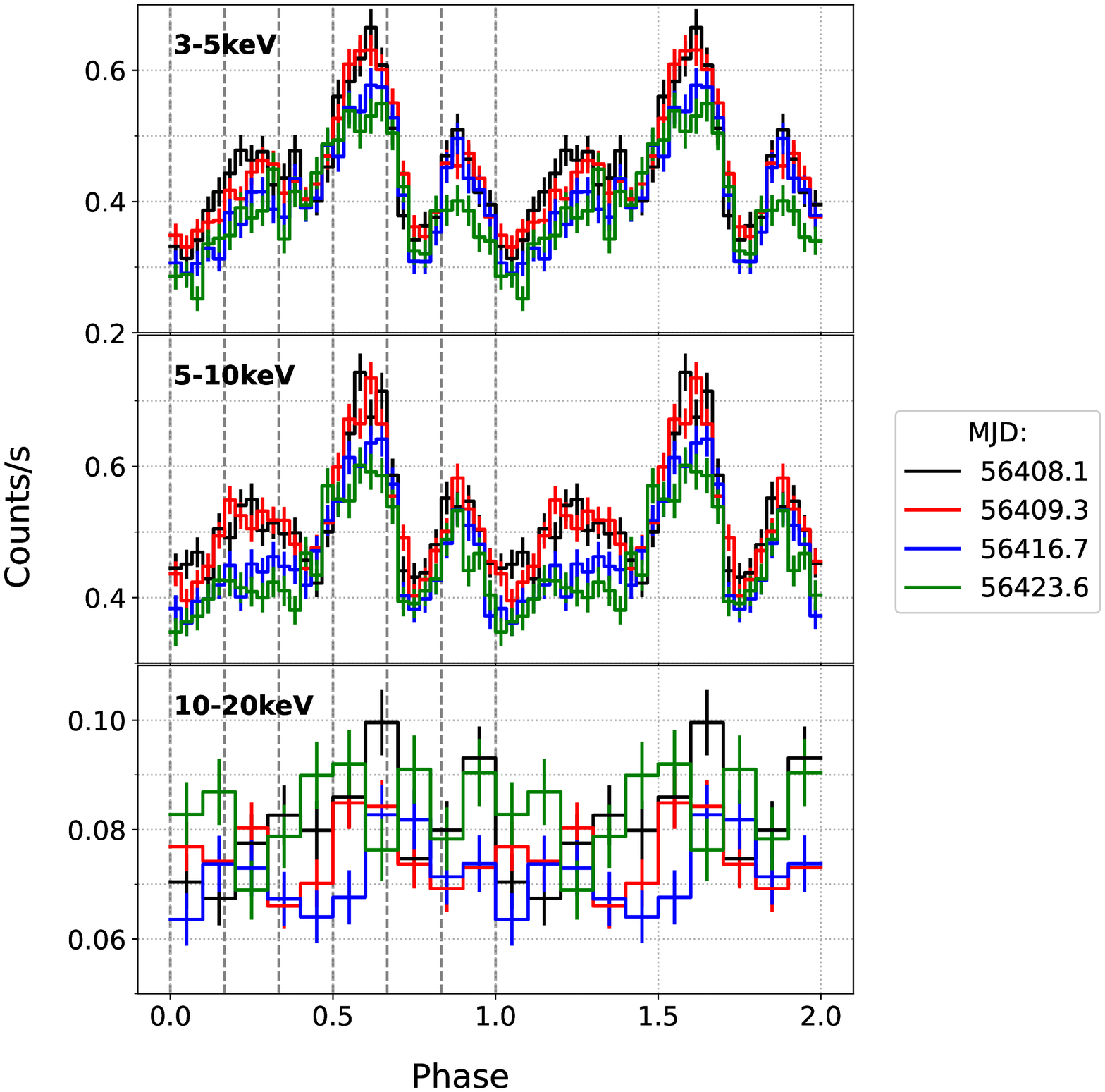}
\includegraphics[width=0.45\textwidth]{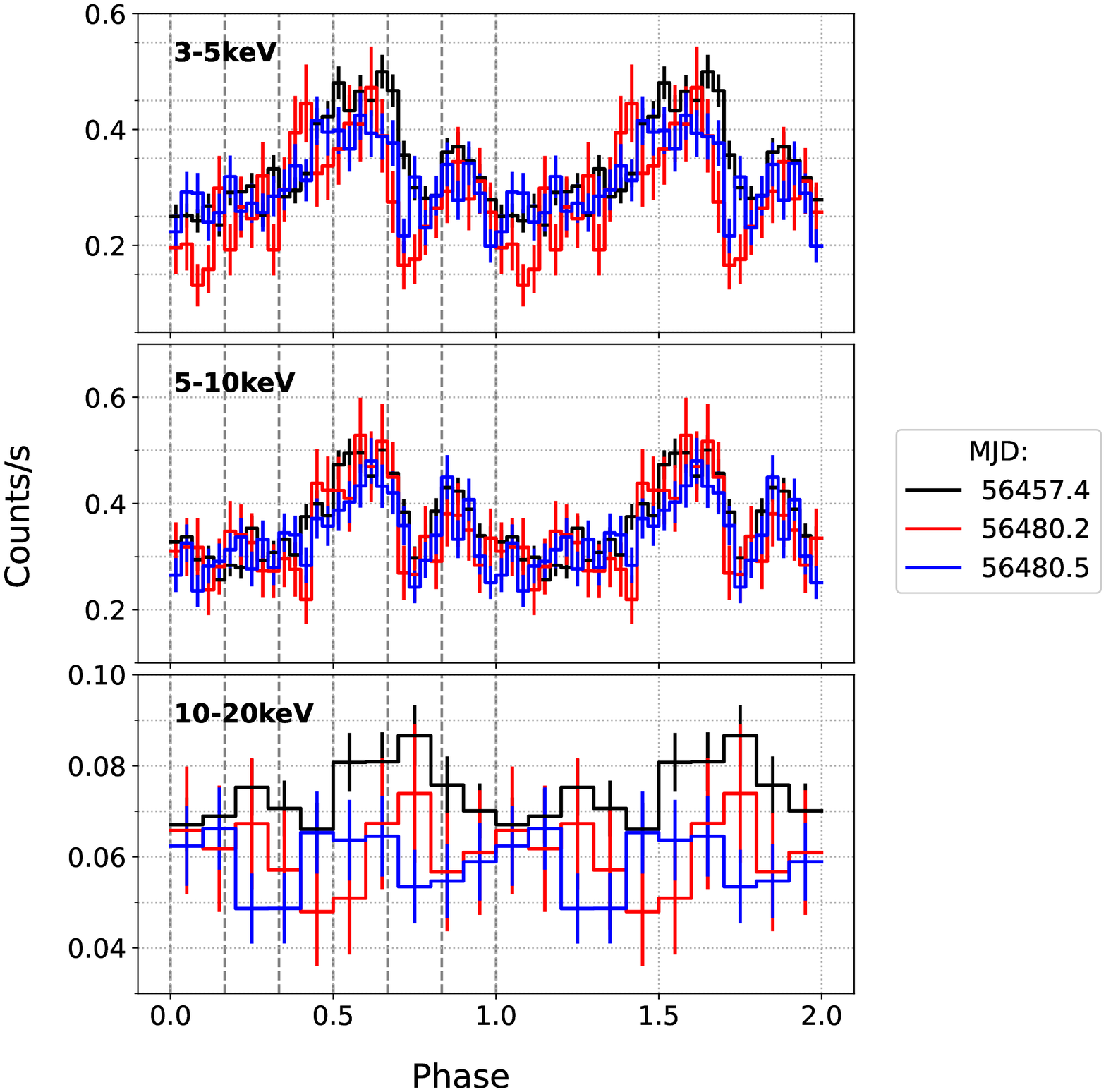}
\caption{Pulse profiles of \sgr\ in different energy bands, 3--5 (upper panels), 5--10 (middle panels), and 10--20~keV (lower panels), in units of the count rate without subtracting the background. Two cycles are presented for clarity. The vertical dashed lines mark the division into phase bins (see Section ''Phase-Resolved Spectroscopy''). }
\label{fig:pp_en}
\end{figure*}

We used the same data set as that in \citet[][see Table 1 in this paper]{Kaspi14} as the initial one. \nus\ has the well-known problem of data contamination by ghost rays, when the detectors are illuminated by emission from sources outside the telescope field of view \citep{Madsen17}. The \fpmb\ data were contaminated in observations 80002013014 and 80002013016 by ghost rays from an unknown source and, therefore, we used only the \fpma\ data for these observations. We also excluded observations 80002013008 and 800020130010, because the bright X-ray source \cxo\ offset from \sgr\ by $24\farcs8$ was observed at this time. In addition, observations 80002013018--80002013024 were excluded from our analysis, because during these observations the X-ray source \ax\ offset from \sgr\ by 87\farcs3 flared up. This source is a low-mass binary that changed its state in the time of its observations with the \nus\ telescope (80002013018 --- the hard state, 80002013020-80002013024 --- the soft one; \citealt{Ponti18}). Since the source \ax\ is brighter than the magnetar \sgr\ by a factor of $\sim10$ and the point spread functions (PSFs) of \ax\ and \sgr\ overlap, the emission component of the source \ax\ that cannot be eliminated may be added to the emission from the magnetar when extracting the spectra. \cite{Kaspi14} estimated the contribution from the source \ax\ to be 3.5\% in a $30\arcsec$ region centered at the magnetar position. Even such a small contribution can distort significantly the results of our phase-resolved spectroscopy for the faint source \sgr\ and, therefore, we decided not to use the observations with the active source \ax\ in this paper. The list of observations used in this paper is given in Table~\ref{tab:obs}.

The photon arrival times were corrected for the Solar system barycenter using the position of the source \sgr\ with the coordinates R.A.=$17^{\rm h}45^{\rm m}40\fs169$, Dec.=$-29\degr00'29\farcs84$ determined by the {\it Chandra} observatory \citep{Rea13}. The light curves with a time resolution of 0.05~s and the spectra were extracted from a circular region of radius $R = 30\arcsec$ centered at the position of \sgr\ (the construction of the background light curve and spectra will be described below) using the {\sc nuproducts} tool, which is a part of the \nus\ Data Analysis Software package ({\sc nustardas V.1.8.0}) built into the {\sc heasoft} software of version 6.27.2. The {\sc caldb} calibration files of version 1.0.2 were used to analyze the data. A direct analysis of the light curves and spectra was performed with two tools, {\sc xronos} of version 5.22 and {\sc xspec} of version 12.10.1 \citep{Arnaud96}, from the {\sc heasoft} package of version 6.27.2.

%*************************************************************
\section*{Timing analysis}
\label{sec:time}

First we obtained the light curves of the source \sgr\ in three energy bands, 3--5, 5--10, and 10--20~keV, for all of the observations from Table~\ref{tab:obs}. To construct the pulse profiles in these energy bands, we used the ephemeris from \citet[][see Table 2 and Section 2.1 in this paper]{Kaspi14}. For each ephemeris we chose the zero epoch in such a way that the minima of the pulse profiles coincided. The light curve for each observation was folded with the corresponding spin period using the {\sc efold} package.

\begin{figure*}[!h]
\centering
\includegraphics[width=0.32\textwidth]{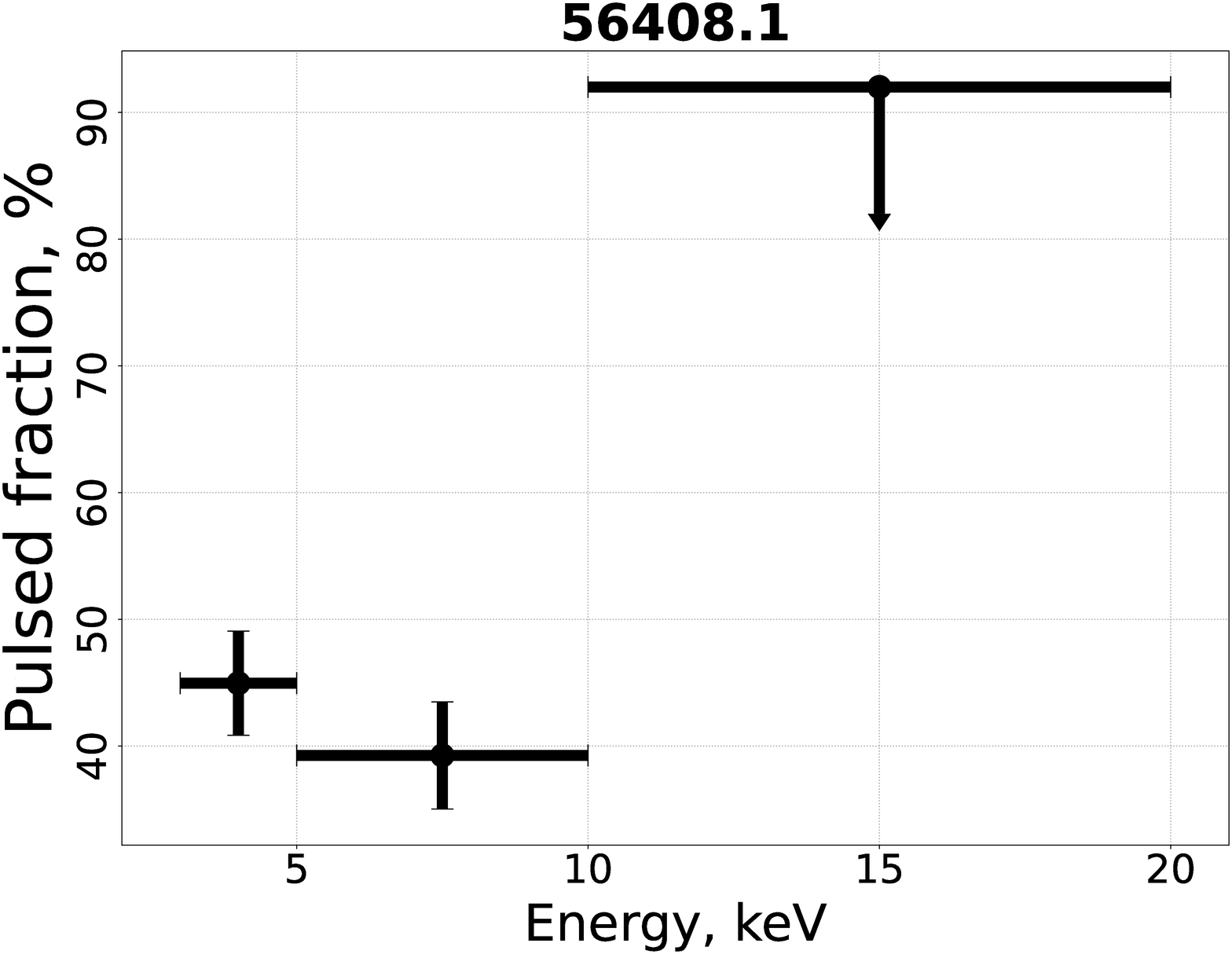}
\includegraphics[width=0.32\textwidth]{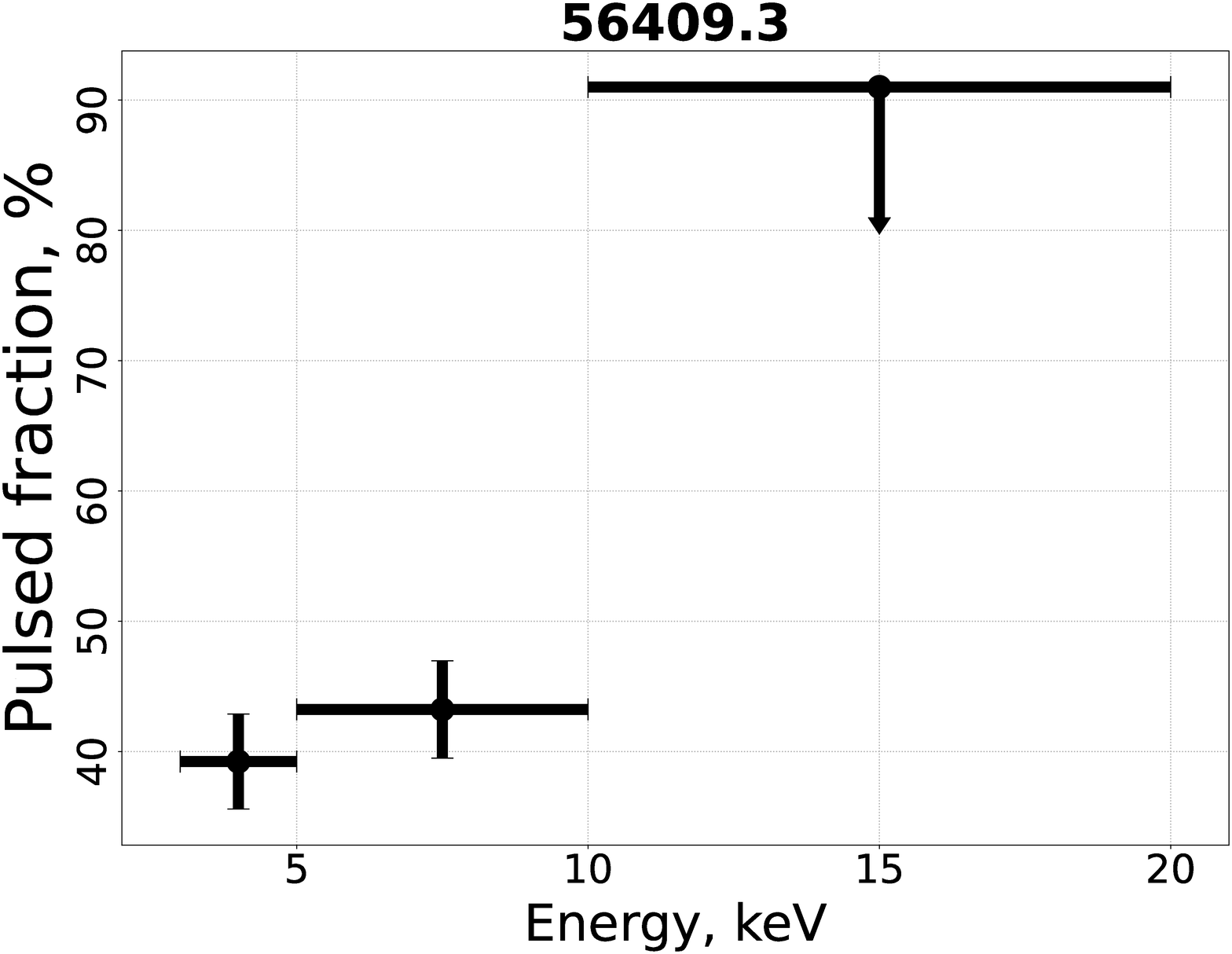}
\includegraphics[width=0.32\textwidth]{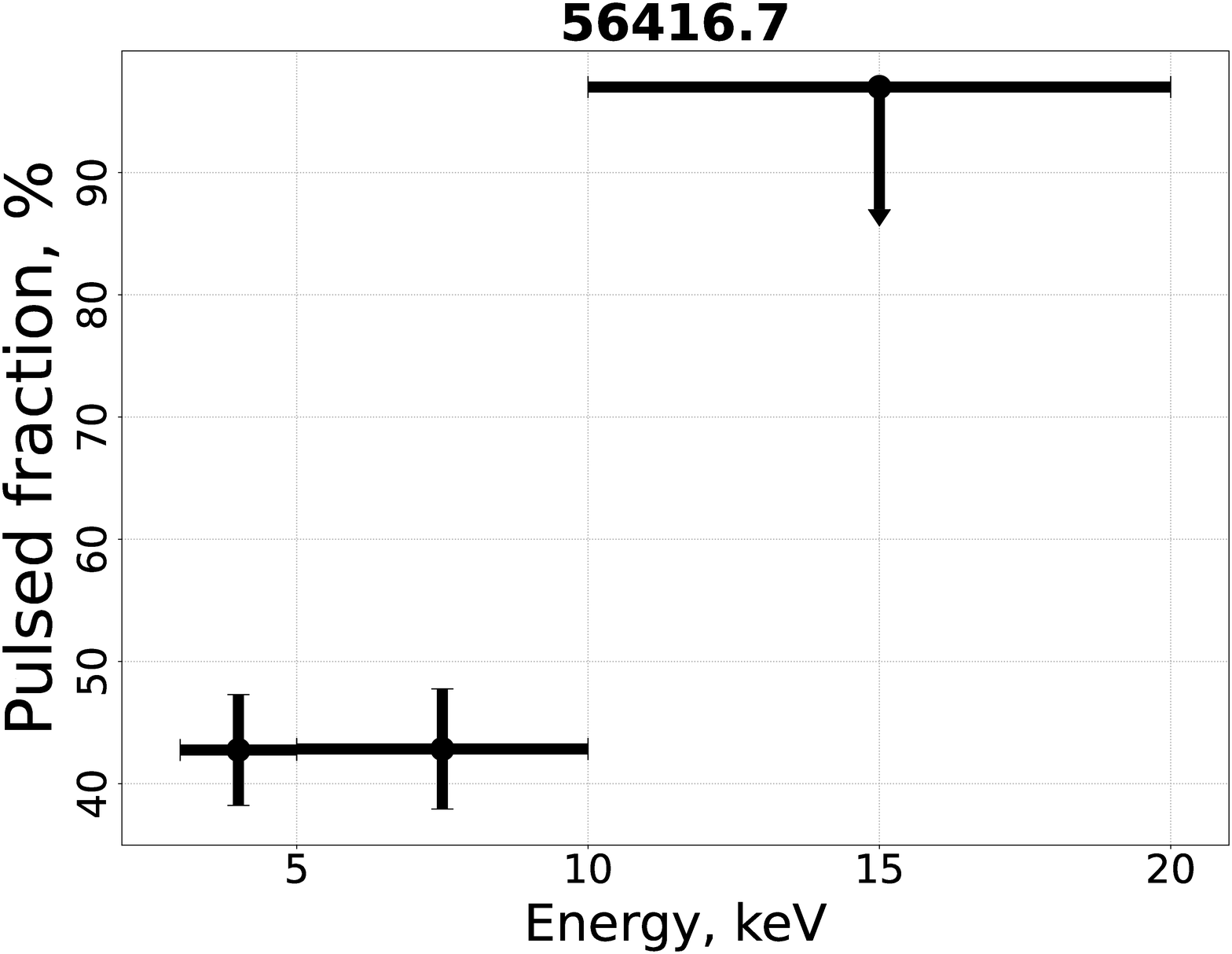}
\includegraphics[width=0.32\textwidth]{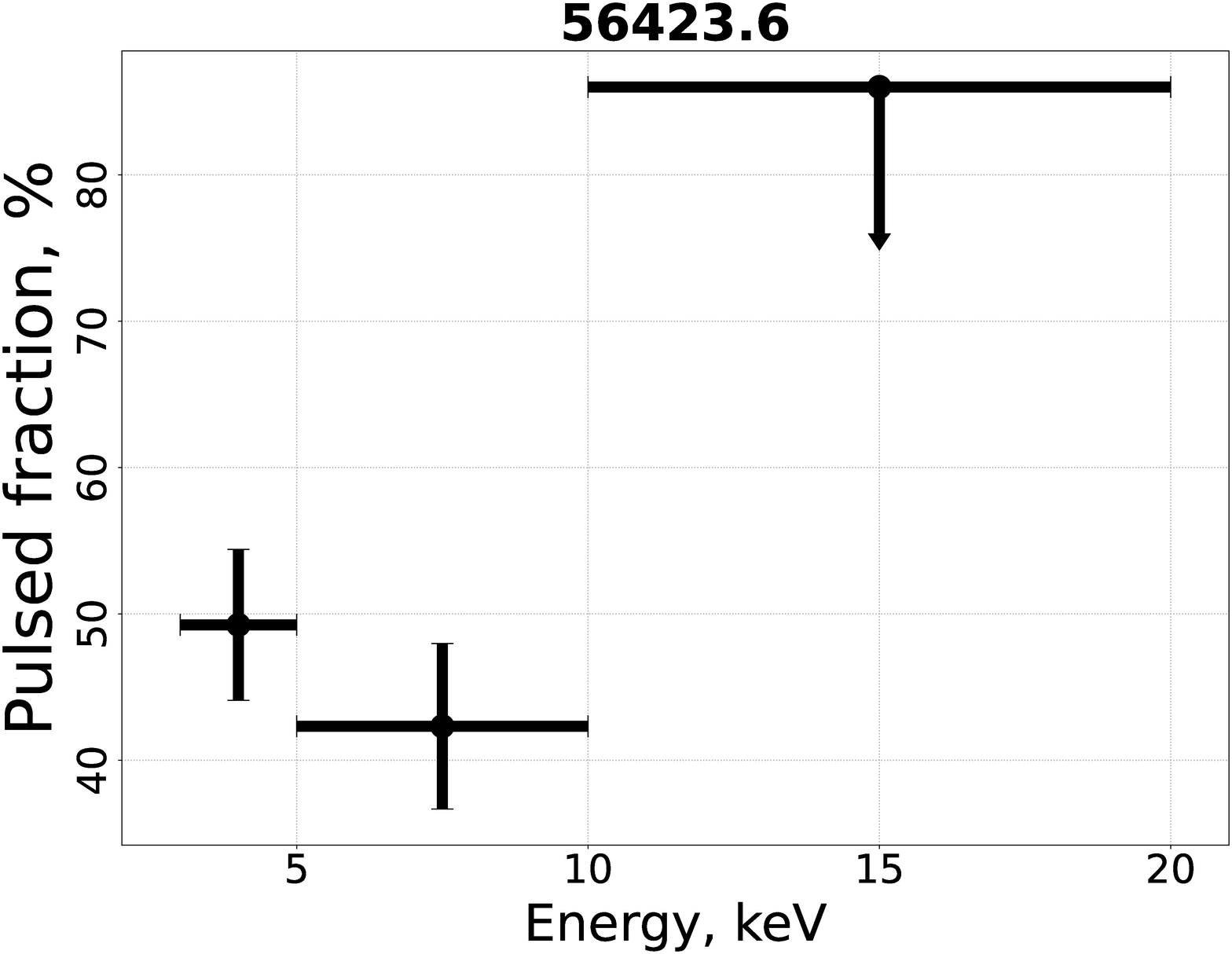}
\includegraphics[width=0.32\textwidth]{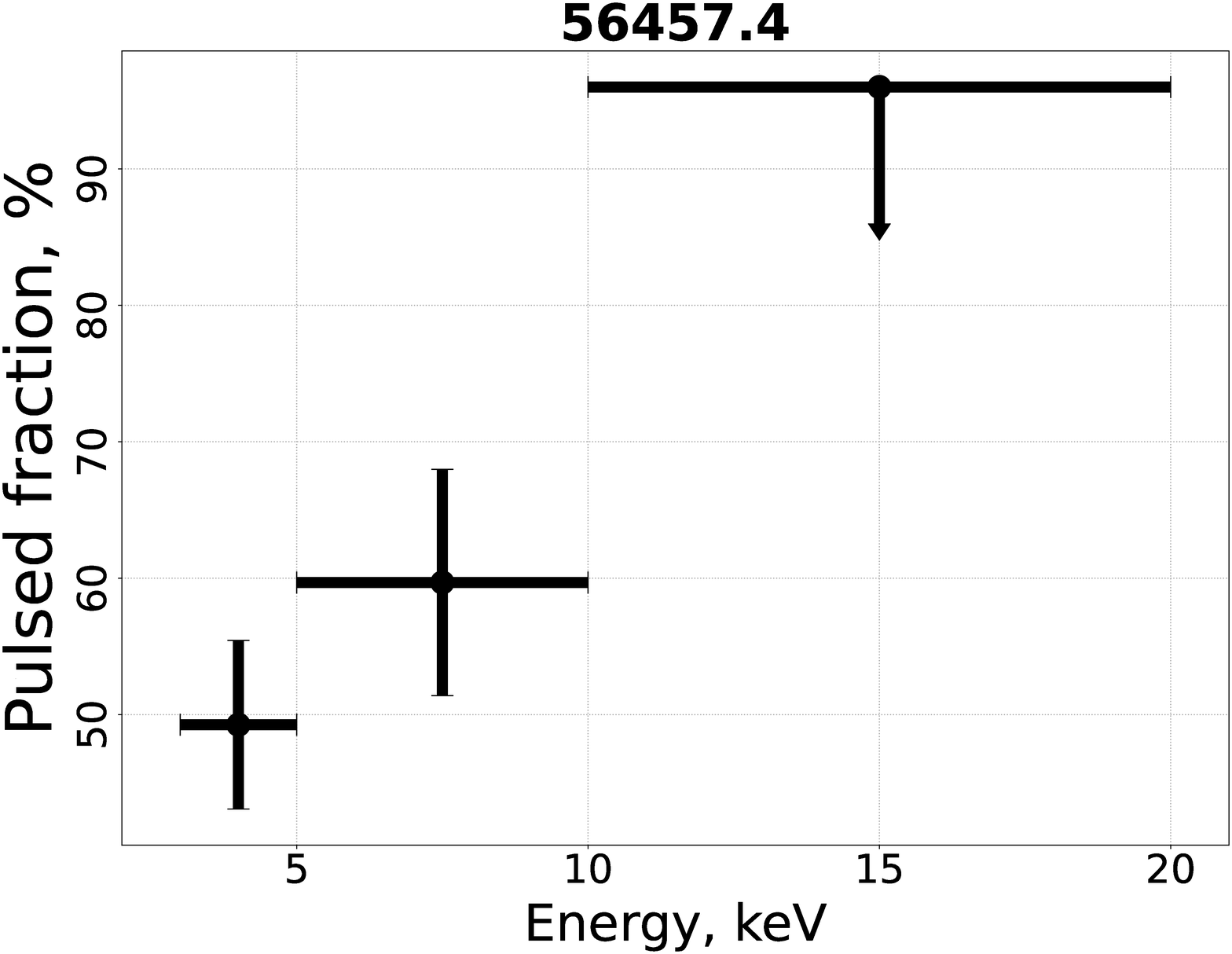}
\includegraphics[width=0.32\textwidth]{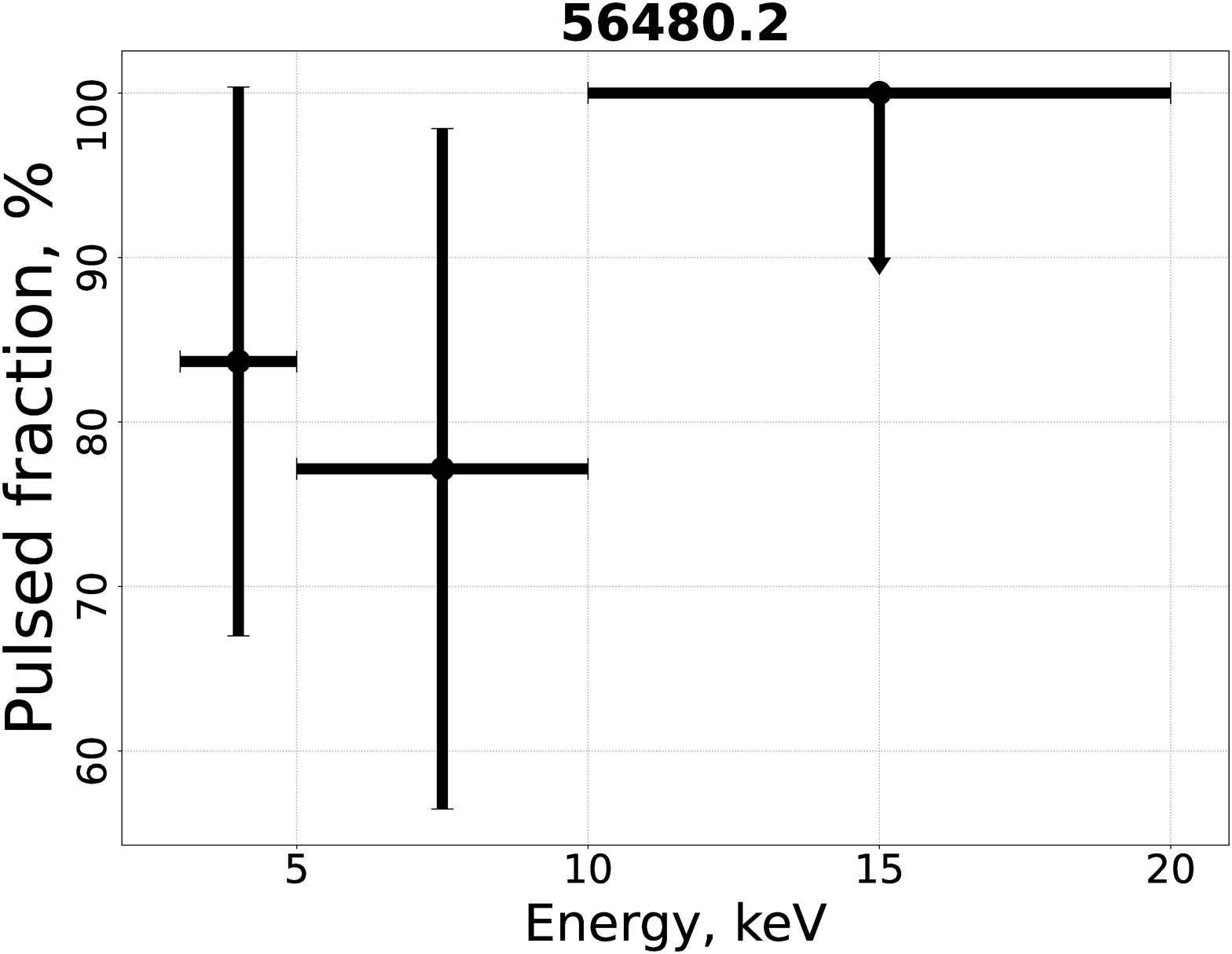}
\includegraphics[width=0.32\textwidth]{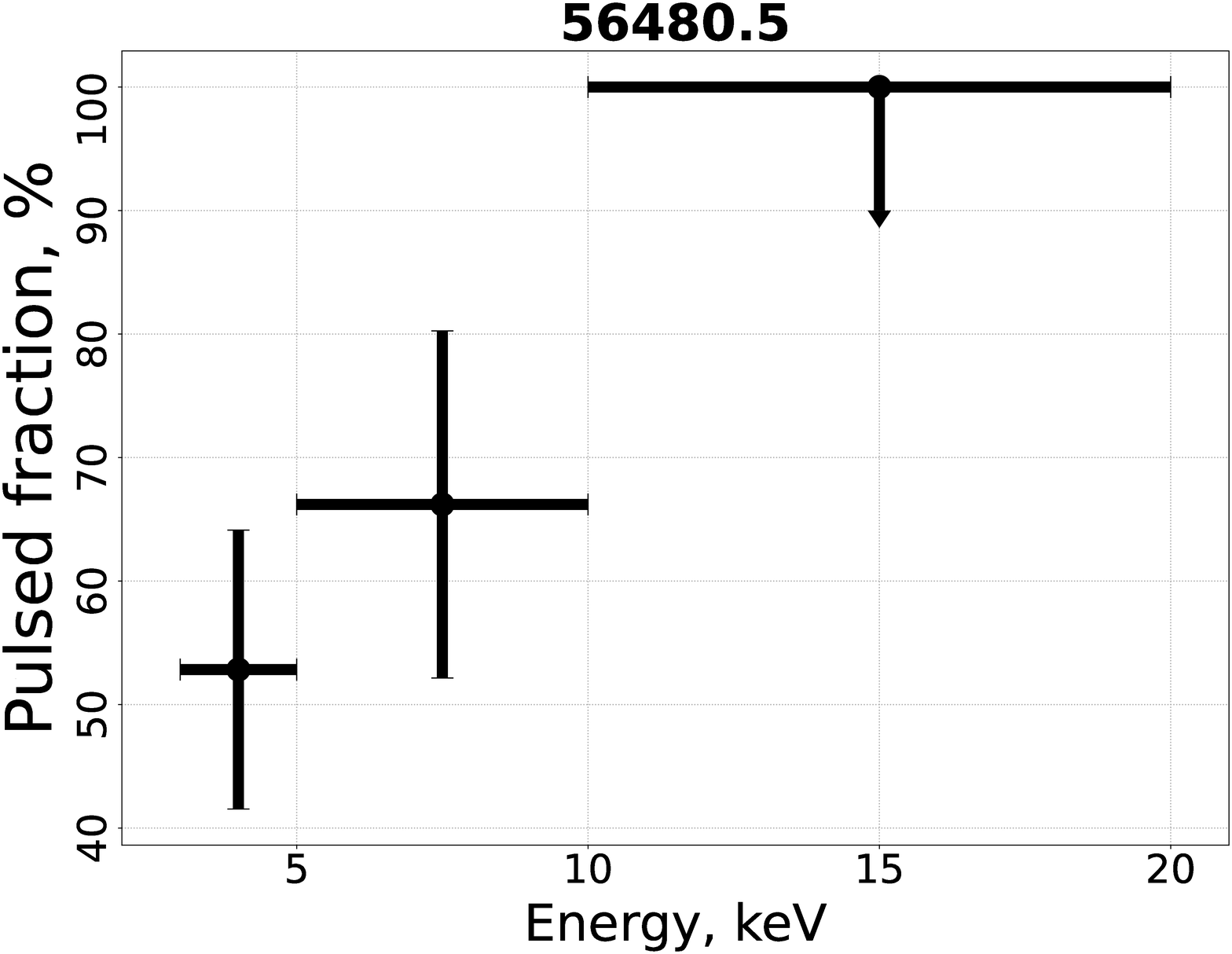}
\caption{Pulsed fraction for seven observations in the 3--5, 5--10, and 10--20~keV energy bands. MJD of each observation is specified in the header of the panels. The arrows indicate the upper limits for the pulsed fraction at a 90\% confidence level.}
\label{fig:pf}
\end{figure*}

The pulse profiles presented in Fig.~\ref{fig:pp_en} have three bright peaks that are clearly visible up to 10~keV. It can be noticed that the intensity drops with time and the first peak is smoothed out. In addition, the intensity of the first peak decreases in the 5--10~keV energy band compared to the peak at energies 3--5~keV. It can be assumed from Fig.~\ref{fig:pp_en} that pulsations may be present at energies above 10~keV. We checked this for the light curves of \sgr\ constructed for the 10--20~keV energy band by two methods, Lomb Scargle \citep{Press89} and the $Z^2_n$ statistic \citep{Buccheri83}, and found no pulsations with upper limits on pulsed fractions of 86--100\%, depending on the observation (see Fig.~\ref{fig:pf} and the explanations below). However, because of the poor statistics at high energies, we cannot unequivocally assert that there are no pulsations in this energy band.

\begin{figure}[!b]
\centering
\includegraphics[width=0.53\textwidth]{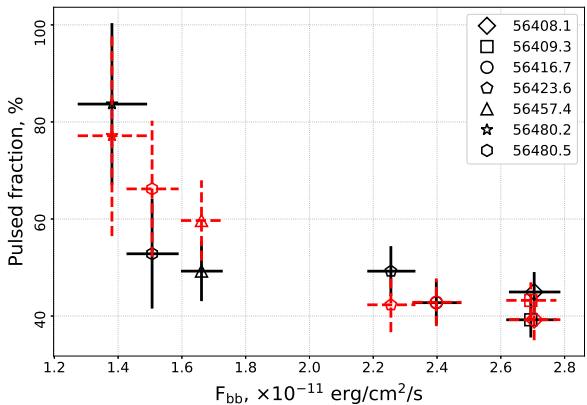}
\caption{Pulsed fraction for energies 3--5 (solid lines) and 5--10~keV (dashed lines) versus thermal {\sc BBrad} flux (see Section ''Spectral Analysis'').}
\label{fig:pf_flux}
\end{figure}

\begin{table*}
\noindent
\begin{center}
\caption{The best-fit parameters when jointly fitting the spectra of the source \sgr\ extracted from a 30\arcsec\ region by the {\sc TBabs*(BBrad+pow)} model.}
\vspace{2mm}
\begin{tabular}{c|c|c|c|c}
\hline
  MJD     & kT, keV       & $R_{BB}$, km                    & $N_{\rm pow}$, $10^{-13}$          & $C_{\rm AB}$           \\
          &               &                                 & erg~s$^{-1}$sm~$^{-2}$~keV$^{-1}$  &            \\
  \hline
  56408.0 & $1.00\pm0.02$ & $1.72\substack{+0.10 \\ -0.09}$ & $1.03\substack{+0.32 \\ -0.26}$ & $0.98\pm0.03$  \\ 
  56409.3 & $0.99\pm0.02$ & $1.74\substack{+0.10 \\ -0.08}$ & $0.84\substack{+0.25 \\ -0.21}$ & $1.00\pm0.04$  \\ 
  56416.7 & $0.97\pm0.02$ & $1.73\substack{+0.10 \\ -0.09}$ & $0.49\substack{+0.21 \\ -0.19}$ & $0.98\pm0.03$  \\ 
  56423.6 & $0.95\pm0.02$ & $1.78\substack{+0.13 \\ -0.11}$ & $1.40\substack{+0.35 \\ -0.30}$ & $0.96\pm0.04$  \\ 
  56457.4 & $0.88\pm0.02$ & $1.85\substack{+0.17 \\ -0.14}$ & $0.94\substack{+0.28 \\ -0.24}$ & $1.01\pm0.05$  \\ 
  56480.2 & $0.93\pm0.06$ & $1.46\substack{+0.28 \\ -0.20}$ & $0.29\substack{+0.46 \\ -0.29}$ & - - -          \\ 
  56480.5 & $0.89\pm0.03$ & $1.72\substack{+0.22 \\ -0.17}$ & $0.10\substack{+0.29 \\ -0.10}$ & - - -          \\ 
  \hline
\end{tabular}\label{tab:fit_all} 
\end{center}
\small{$C_{\rm AB}$ is the cross-normalization constant between the \fpma\ and \fpmb\ data determined for each observation individually. The power-law normalization $N_{\rm pow}$ is given for an energy of 10~keV.}
\end{table*}

To estimate the pulsed fraction\footnote{The pulsed fraction was determined from the formula $PF=(I_{max}-I_{min})/(I_{max}+I_{min})$, where $I_{max}$ and $I_{min}$ are the intensities at the pulse maximum and minimum.}, we first subtracted the background from the light curves of \sgr. It should be noted that \sgr\ is located in the immediate vicinity of the SMBH \sgra\ and the \nus\ observatory cannot resolve these two objects, which makes it much more difficult to obtain the correct background estimate. In addition, the background emission near the Galactic center is spatially inhomogeneous \citep[see, for e.g.][]{Perez15}, which does not allow a region free from point sources and far from the magnetar \sgr\ to be used to estimate the background either. Therefore, as in \cite{Kaspi14}, we used the previous \nus\ observation 30001002003 of the Galactic center region, during which the magnetar \sgr\ was not yet in its active state, to estimate the background. We consider whether this background estimation method is appropriate in Section ''Spectral Analysis''. The background light curve was extracted for different energy bands from the same sky region as that used to extract the information about the \sgr\ emission. The mean background count rates for the 3--5, 5--10, and 10--20~keV energy bands were found to be $0.097\pm0.002$, $0.186\pm0.002$, $0.060\pm0.001$~counts~s$^{-1}$, respectively. These values were subtracted from the source light curves constructed for each observation. The derived pulsed fractions for the 3--5 and 5--10~keV energy bands (Fig.~\ref{fig:pf}) are $\sim40-50\%$, consistent with the results obtained for the soft X-ray band from {\it Chandra} and {\it XMM-Newton} data \citep{Coti15,Coti17}. It can also be noted that the pulsed fractions for energies 3--5 and 5--10~keV are in a good agreement between themselves within one observation.

Such high pulsed fractions may point to an asymmetric arrangement of two opposite thermal emission regions \citep{Beloborodov02}. However, using the 2016 data, when the magnetar pulse profile underwent significant changes, \cite{Hu19} suggested that two approximately symmetric opposite emission regions, whose intensities differ by more than a factor of $\sim3$, are observed for \sgr. Note also that \cite{Hu19} used a slightly different definition of the pulsed fraction that systematically gives lower values than ours.

Markov-chain Monte Carlo simulations \citep{Sitter01} were performed to estimate the upper limit on the pulsed fraction in the 10--20~keV energy band. The count rate in each phase bin was assumed to distributed normally with the values obtained using the {\sc efold} package. The background count rate determined above was first subtracted from the count rate in each bin. The a~priori distribution of count rates from the pulsar at each phase was assumed to be distributed uniformly in the range 0.0--0.2~count~s$^{-1}$, which is definitely above the recorded count rate in phase bins. The upper limit on the pulsed fraction was taken to be equal to the 90\% quantile for the pulsed fraction of the a posteriori distribution of count rates in phase bins. The derived values are specified in Fig.~\ref{fig:pf} as upper limits.

We also plotted the pulsed fraction against the thermal {\sc BBrad} flux (Fig.~\ref{fig:pf_flux}) gathered from the same circular 30\arcsec\ region as that used for the light curves. It can be noticed from Fig.~\ref{fig:pf_flux} that the pulsed fraction derived for two energy bands, 3--5 and 5--10~keV, increases with decreasing flux.

To describe the dependence of the pulsed fraction on energy and flux, we used several models:
\begin{itemize}
  \item the pulsed fraction is constant, $PF=C$;
  \item the pulsed fraction depends linearly on the luminosity, $PF=C+F_c*F$;
  \item the pulsed fraction depends linearly on the luminosity and energy, $PF=C+F_c*F+E_c*E$,
\end{itemize}
where $C$ is the constant component of the pulsed fraction, $F_c$ is the linear correlation coefficient of the pulsed fraction and flux $F$ in the 3--20~keV energy band, and $E_c$ is the linear correlation coefficient of the pulsed fraction and energy $E$.

The pulsed fractions were fitted by these models. As a result, we obtained the following values of the data likelihood $\chi^2/(d.o.f.)$: 36.0/20, 20.1/19 and 19.9/18 for each of the models, respectively. The relative significances of the models were checked using the criterion for cross-checking the samples of parameters obtained in our Markov-chain Monte Carlo (MCMC) simulations based on leave-one-out algorithms \citep{Vehtari14}. The criterion showed that the model of a linear dependence of the pulsed function on luminosity is much more probable than the model with a constant pulsed fraction ($P_{\rm fc}/P{\rm nc} = 0.978/0.022$), where $P_{\rm fc}$ is the relative probability of the model with a linear dependence of the pulsed fraction on flux and $P_{\rm nc}$ is the relative probability of the model with a constant pulsed fraction. The model that additionally includes a linear dependence of the pulsed fraction on energy and luminosity with a relative probability $P_{\rm ec \& fc}$ turned out to be statistically comparable to the model in which there is no dependence on energy: $P_{\rm ec \& fc}/P_{\rm fc} = 0.41/0.59$.

\begin{figure}
\centering
\includegraphics[width=0.5\textwidth]{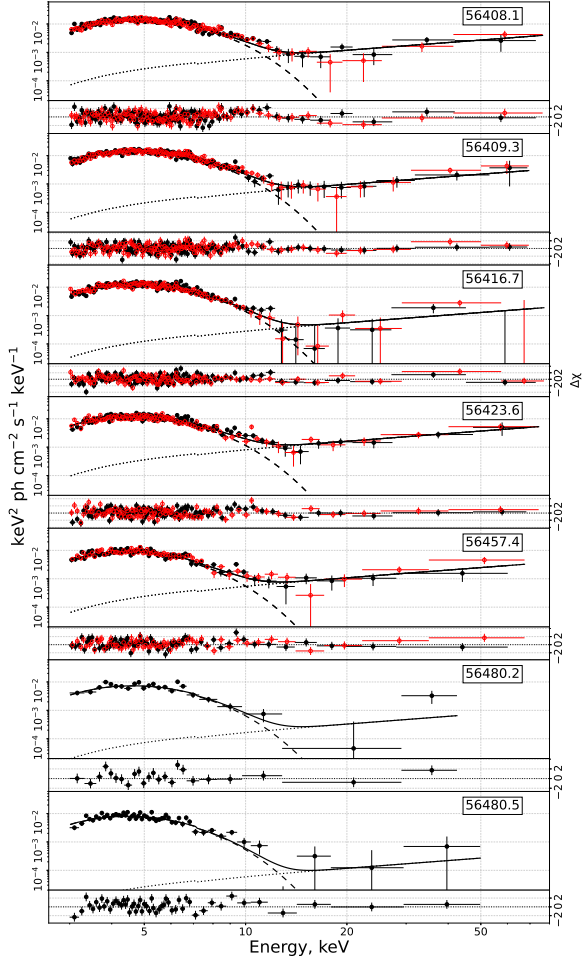}
\caption{The \sgr\ spectrum measured in different observations with MJD of their beginning shown in the figure. The black and red colors mark the \fpma\ and \fpmb\ data, respectively. The solid, dashed, and dotted lines indicate the total, blackbody, and power-law models, respectively.}
\label{fig:fit_spec}
\end{figure}

\section*{Spectral analysis}
\label{sec:spec}

Phase-resolved spectroscopy of pulsating sources is an important method for studying the emission generation mechanisms. The sky region containing the magnetar \sgr\ is very densely populated, making it much more difficult to analyze the data. Therefore, before turning to our phase-resolved spectroscopy of the magnetar, we obtained the average spectra for each observation and compared them with those obtained previously by other authors to make sure that the procedure of spectral analysis used by us is correct.

\subsection*{Average spectra}
\label{sec:Total-spectra}

The source spectra and light curves were extracted from the same sky region. As the background we used the previous observation 30001002003 and the same region that were previously chosen to determine the background for the light curves. Note that \cite{Kaspi14} also considered a different, more complex method of estimating the background emission for each observation and showed that the results in both cases agree between themselves. Therefore, we chose the most optimal approach described above for our estimates.

Each spectrum was grouped using the {\sc grppha} tool, which is a part of the {\sc heasoft} software, with a minimum number of counts per bin equal to 25. A number of authors \citep[see, for e.g.][]{Mori13,Kaspi14,Coti17} showed that the source spectra can be best fitted by a combination of blackbody radiation ({\sc BBrad}) and a power-law ({\sc pow}) with absorption $N_{\rm H}$. To describe the latter, we used the {\sc TBabs} model with the abundance from \cite{Wilms2000} and the absorption cross sections from \cite{Verner96}.

\begin{figure*}[!h]
\centering
\includegraphics[width=0.8\textwidth]{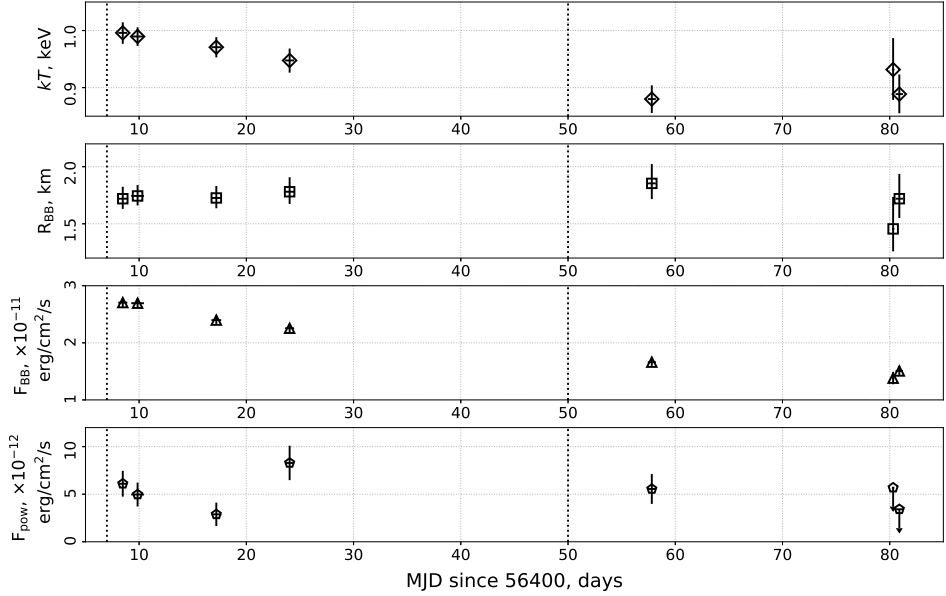}
\caption{Dependences of the parameters of the average spectra $kT$ ,$R_{\rm BB}$, $F_{\rm BB}$, and $F_{\rm pow}$ on the MJD epoch taken for the middle of the observation shown on the graphs from top to bottom, respectively. The black dashed lines mark the times of the outbursts detected with \swift/BAT on April 25 \citep{Kennea13a} and June 7 \citep{Kennea13b}.}
\label{fig:fit_all}
\end{figure*}

The spectra for all observations were fitted jointly by assuming that the absorption column density $N_{\rm H}$ and the photon index $\Gamma$ did not change from observation to observation. It should be noted that at the initial stage for each pair of \fpma\ and \fpmb\ spectra referring to one observation we determined the cross-normalization parameters $C_{\rm AB}$ of the \fpmb\ spectra with respect to the \fpma\ spectra that were fixed while fitting all spectra (for accurate values, see Table~\ref{tab:fit_all}). The temperature $kT$ and radius $R_{\rm BB}$ of the emitting region referring to the {\sc BBrad} model and the power-law normalization $N_{\rm pow}$ were related within one observation for both modules, but were free relative to different observations. As a result of fitting by the {\sc TBabs*(BBrad+pow)} model, we obtained the best-fit values of $N_{\rm H}=(11.5\pm0.8)\times10^{22}$~cm$^{-2}$ and $\Gamma=1.11^{+0.26}_{-0.24}$ with the reduced value of $\chi^2_{\rm red}/d.o.f.=1.05/2229$ (Fig.~\ref{fig:fit_spec}). The best-fit values of the changing parameters are given in Table~\ref{tab:fit_all}.

\begin{figure*}[]
\centering
\includegraphics[width=0.31\textwidth]{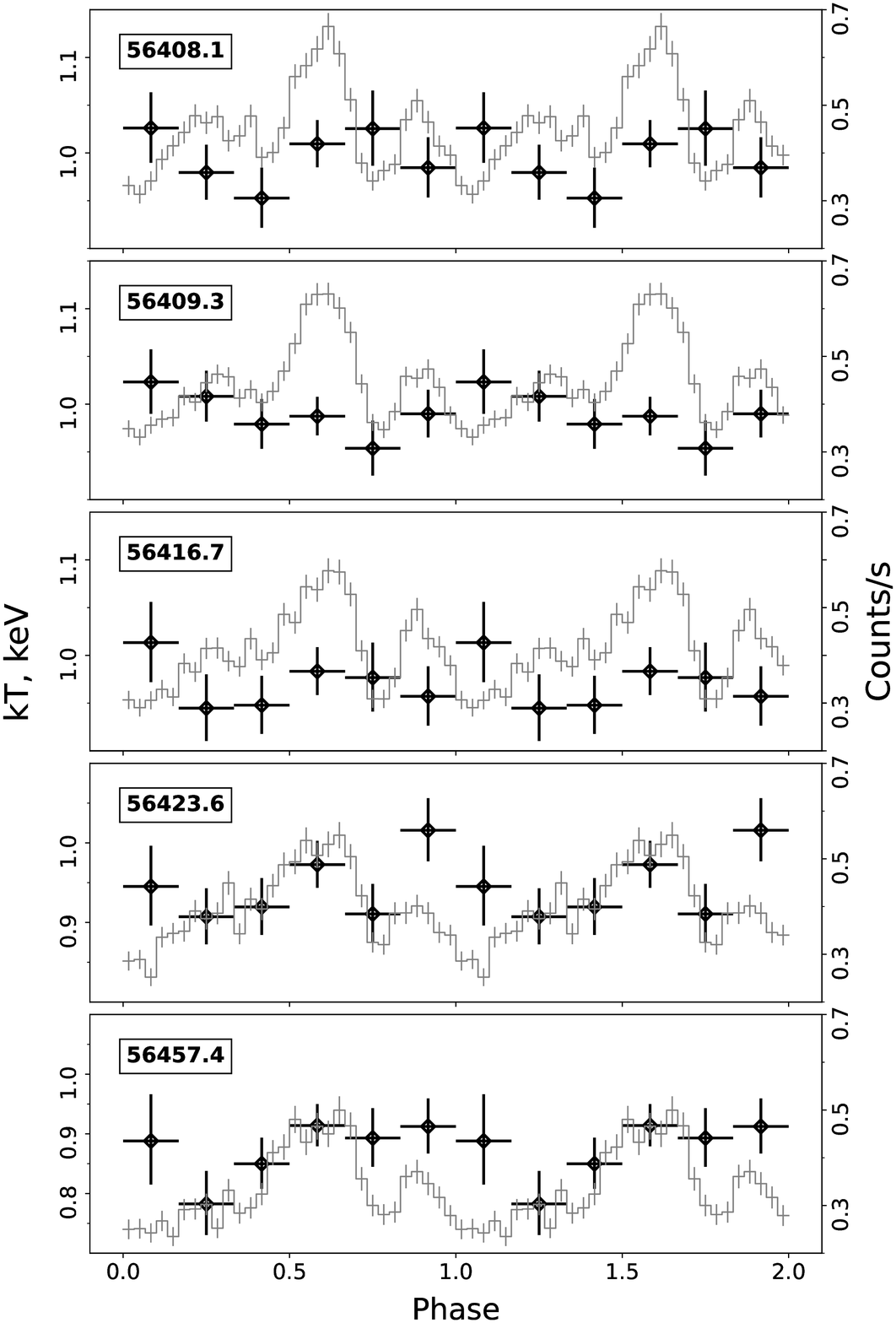}
\includegraphics[width=0.31\textwidth]{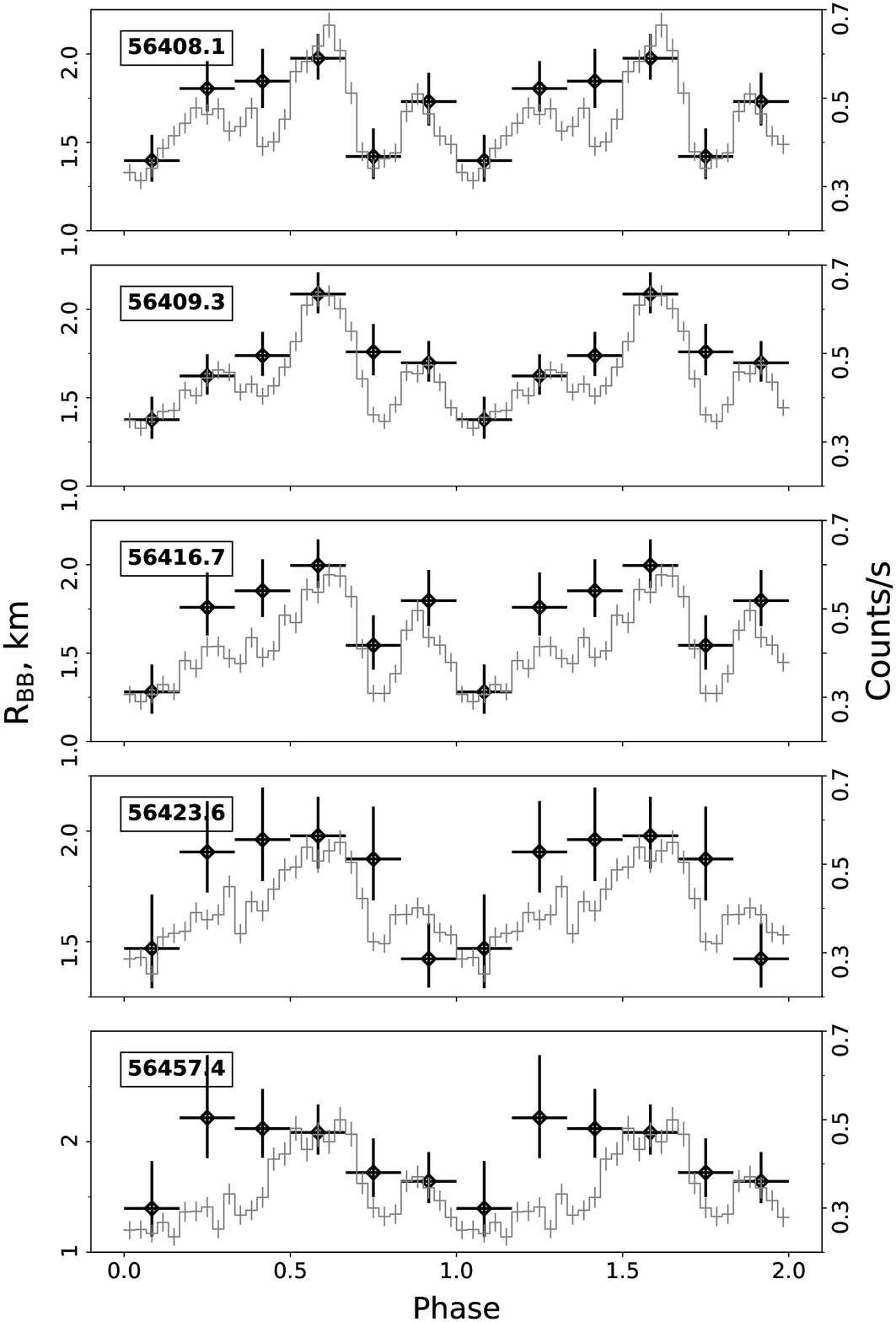}
\includegraphics[width=0.31\textwidth]{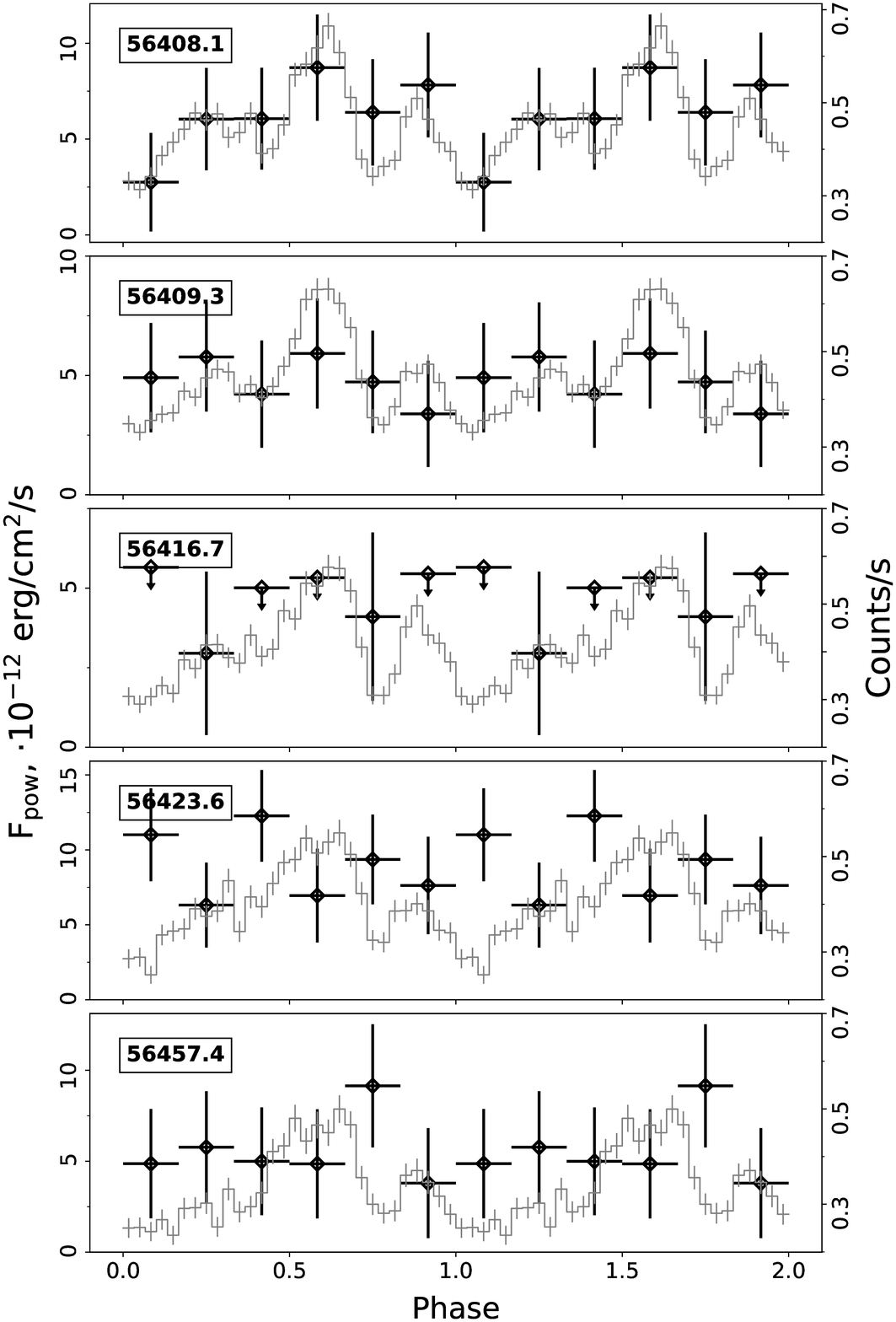}
\caption{Spectral parameters as a function of pulse phase and observation (from top to bottom). The gray color and the right scale represent the corresponding pulse profiles in the 3--5~keV energy band. Two cycles are presented for clarity.}
\label{fig:phased}
\end{figure*}

Figure~\ref{fig:fit_all} shows the time evolution of the parameters $kT$, $R_{\rm BB}$, and the absorbed fluxes calculated using the {\sc cflux} command for the thermal and non-thermal components of the model that were designated as $F_{\rm BB}$ and $F_{\rm pow}$, respectively. A drop in temperature $kT$ by $\sim10\%$ and in flux $F_{\rm BB}$ by $\sim40\%$ is clearly seen on a time scale $\sim80$~days. At the same time, the parameter $R_{\rm BB}$ remains almost constant, within the measurement error limits, on this time scale, although on a longer scale of observations the decrease in $R_{\rm BB}$ becomes noticeable \citep[see][]{Coti15,Coti17,Rea20}. This trend may already begins to have an effect at the end of the time interval under study as well (Fig.~\ref{fig:fit_all}). However, the large errors in $R_{\rm BB}$ in the last two observations do not allow an unequivocal conclusion to be reached. The latter implies that on a time scale $\sim80$~days we do not observe any change in the area of the emitting region A with luminosity $L_{\rm BB}$ predicted by the model of an untwisting neutron star magnetosphere \cite{Beloborodov09} in the form $A\sim L^{1/2}_{\rm BB}$, which is the most suitable model to explain such a slow decay of the magnetar emission \citep{Mori13,Kaspi14,Coti15,Coti17}. Note that a possible decrease in the radius of the emitting region can explain the increase in the pulsed fraction with decreasing flux \citep[see, for e.g.][]{Ozel02}.

On the whole, the results are in a good agreement with those from \cite{Kaspi14}. Note that our estimate of the absorption column density, $N_{\rm H}=(11.5\pm0.8)\times10^{22}$~cm$^{-2}$, is slightly smaller than the value obtained by these authors, $N_{\rm H}=(13.5\pm0.5)\times10^{22}$~cm$^{-2}$, although they are in agreement within a level better than $2\sigma$. This may be because \citep{Kaspi14} used a larger number of observations that we excluded from our analysis (see above). Furthermore, our estimates of the temperature $kT$ of the emitting region are slightly higher, while the estimates of the radius $R_{\rm BB}$ are lower than the estimates of these parameters from {\it Chandra} data \citep{Coti15,Coti17}. This may be because the operating energy band of the {\it Chandra} observatory is softer (0.3--8~keV), and these authors did not use the power-law component in the spectral magnetar emission model.

Thus, we tested the validity of the data reduction and may turn to our phase-resolved spectroscopy.

\subsection*{Phase-Resolved Spectroscopy}
\label{sec:Phase-resolved}

The observed pulse profile was divided into six equal phases in such a way that for each maximum and minimum there was one phase bin (Fig.~\ref{fig:pp_en}). The source spectrum in each phase bin was fitted by the {\sc TBabs*(BBrad+pow)} model that was applied to the average spectra, with the absorption column density and the photon index having been fixed at $N_{\rm H}=11.5\times10^{22}$~cm$^{-2}$ and $\Gamma=1.11$ determined previously. Observations 80002013014 and 80002013016, for which only the \fpma\ data were available, do not have sufficient statistics for a high-quality fit to the phase spectra (the number of degrees of freedom, d.o.f.~$\sim10-40$, is considerably smaller than that in the remaining observations, d.o.f~$\gtrsim 100$) and, therefore, we excluded them from our analysis. The results obtained are presented in Fig.~\ref{fig:phased}. To understand how much a specific parameter changes with phase, we fitted each set of values for all observations by a constant. Thus, we can say that the temperature kT of the emitting region hardly changes with phase ($\chi^2_{red}\leq1$ for the first three observations and $\chi^2_{red}\approx2$ for the last two observations 80002013010 and 80002013012); in contrast, its radius $R_{\rm BB}$ changes with pulse phase more significantly ($\chi^2_{red} > 2$ for all observations, except 80002013012, where $\chi^2_{red}=1.55$). It is visually noticeable that that the $R_{\rm BB}$ variations closely follow the shape of the pulse profile, which can be justified by the visual geometry (i.e., the largest and smallest area responsible for the generation of thermal emission is visible at the maxima and minima of the pulse profile, respectively). Taking into account the above dependence of the pulsed fraction on flux, one might expect similar dependences for the parameters of the thermal component as well. However, our analysis did not reveal significant changes in the variability amplitude of the parameters $kT$ and $R_{\rm BB}$ with thermal flux $F_{\rm BB}$, which may be due to significant errors in their values.

A significant increase in $R_{\rm BB}$ in the second phase bin of the last observation is worth noting. Interestingly, in the same phase bin in the last observation the first peak of the pulse profile virtually disappears. Furthermore, a reduced $kT$ can be noticed in the same region. The latter may be due to both physical factors and possible anti-correlation of model parameters. However, we cannot reach any unequivocal conclusions because of the rather poor statistics.

The non-thermal flux $F_{\rm pow}$ hardly changes with phase, which may suggest the generation of non-thermal emission in other regions with respect to the hot spots or insufficient statistics for the detection of its variability.

\section*{Conclusions}

The magnetar \sgr\ has been an object of a large number of observations from the beginning of its activity in April 2013. In particular, the program of observations with the \nus\ observatory provided a good opportunity to study the hard X-ray emission from this object. These observations allowed timing and spectral analyses based on \nus\ data to be performed (see the references above in the text). However, the phase-resolved spectroscopy results were briefly presented only for the first observation \citep{Mori13}.

In this paper, based on data from the \nus\ observatory, for the first time we have performed detailed phase-resolved spectroscopy for the magnetar \sgr\ in a wide energy band for states with different intensities of the source. As a result, we found significant changes in the apparent sizes of the region responsible for the thermal emission correlating with the pulse profile in the 3--5~keV energy band. The temperature of the emitting region remains fairly stable on pulse, while decreasing, on average, with decreasing intensity of the source. Unfortunately, the magnetar \sgr\ is too faint to perform detailed phase-resolved spectroscopy for its non-thermal emission. We found no significant changes in the total flux of the power-law component with a fixed photon index $\Gamma=1.11$. However, the available statistics does not allow us to unequivocally assert that the non-thermal component indeed does not pulsate.

Apart from phase-resolved spectroscopy, we estimated the pulsed fraction for two energy bands, 3--5 and 5--10~keV, to be $\sim40-50\%$. We found evidence for a significant increase in the pulsed fraction with decreasing flux from the magnetar, while the dependence on energy is not determined at a statistically significant level presumably due to a possible decrease in the radius of the thermal emission generation region.

\section*{Funding}

This study was supported by grant no. 14.W03.31.0021 from the Government of the Russian Federation. E.A. Kuznetsova also thanks the Russian Foundation for Basic Research (project. 19-32-90283) for its partial support of this work with regard to the studies of extended emission to obtain proper background estimates.

\bibliographystyle{mnras}
\bibliography{biblio} 

\end{document}